\documentclass[sigconf,anonymous=False]{acmart}

\usepackage{multirow}
\usepackage{amsmath}
\usepackage{subfig} 
\usepackage[flushleft]{threeparttable}
\usepackage{makecell} 
\usepackage{enumitem} 

\usepackage{array}
\newcommand{\PreserveBackslash}[1]{\let\temp=\\#1\let\\=\temp}
\newcolumntype{C}[1]{>{\PreserveBackslash\centering}p{#1}}
\newcolumntype{R}[1]{>{\PreserveBackslash\raggedleft}p{#1}}
\newcolumntype{L}[1]{>{\PreserveBackslash\raggedright}p{#1}}

\makeatletter
\newcommand\footnoteref[1]{\protected@xdef\@thefnmark{\ref{#1}}\@footnotemark}
\makeatother

\newcommand\blfootnote[1]{%
\begingroup 
\renewcommand\thefootnote{}\footnote{#1}%
\addtocounter{footnote}{-1}%
\endgroup 
}

\AtBeginDocument{%
  }

\begin{document}
\fancyhead{}

\title{Disentangled Modeling of Domain and Relevance for \\ Adaptable Dense Retrieval}


\author{Jingtao Zhan$^{1}$, Qingyao Ai$^{1}$, Yiqun Liu$^{1\star}$, Jiaxin Mao$^{2}$, Xiaohui Xie$^{1}$, Min Zhang$^{1}$, Shaoping Ma$^{1}$}
\affiliation{%
  \institution{${1}$ Department of Computer Science and Technology, Beijing National Research Center for Information Science and Technology, Tsinghua University, Beijing 100084, China}
  \country{}
}
  
\affiliation{%
  \institution{${2}$ Beijing Key Laboratory of Big Data Management and Analysis Methods, Gaoling School of Artificial Intelligence, \\ Renmin University of China, Beijing 100872, China}
  \country{}
}

\affiliation{%
	\institution{\{jingtaozhan, aiqingyao, maojiaxin\}@gmail.com, xiexiaohui@mail.tsinghua.edu.cn, \\ \{yiqunliu, z-m, msp\}@tsinghua.edu.cn}
	\country{}
}



\renewcommand{\shortauthors}{Zhan, et al.}

\begin{abstract}
Recent advance in Dense Retrieval~(DR) techniques has significantly improved the effectiveness of first-stage retrieval. Trained with large-scale supervised data, DR models can encode queries and documents into a low-dimensional dense space and conduct effective semantic matching. However, previous studies have shown that the effectiveness of DR models would drop by a large margin when the trained DR models are adopted in a target domain that is different from the domain of the labeled data. One of the possible reasons is that the DR model has never seen the target corpus and thus might be incapable of mitigating the difference between the training and target domains. In practice, unfortunately, training a DR model for each target domain to avoid domain shift is often a difficult task as it requires additional time, storage, and domain-specific data labeling, which are not always available. To address this problem, in this paper, we propose a novel DR framework named Disentangled Dense Retrieval~(DDR) to support effective and flexible domain adaptation for DR models. DDR consists of a Relevance Estimation Module~(REM) for modeling domain-invariant matching patterns and several Domain Adaption Modules~(DAMs) for modeling domain-specific features of multiple target corpora. By making the REM and DAMs disentangled, DDR enables a flexible training paradigm in which REM is trained with supervision once and DAMs are trained with unsupervised data. Comprehensive experiments in different domains and languages show that DDR significantly improves ranking performance compared to strong DR baselines and substantially outperforms traditional retrieval methods in most scenarios.
\end{abstract}

\ccsdesc[500]{Information systems~Retrieval models and ranking}

\keywords{dense retrieval, disentangled  modeling, domain adaption}

\maketitle

\makeatletter
\if@ACM@anonymous
\else
	\blfootnote{$^\star$Corresponding author}
\fi
\makeatother

\section{Introduction}

Dense Retrieval~(DR) has become an important and popular technique for first-stage retrieval~\cite{xiong2021approximate, zhan2022learning, gao2021unsupervised}. 
With a sufficient amount of supervised data, DR models can map queries and documents into a latent embedding space and estimate relevance based on embedding similarities, which significantly alleviates the vocabulary mismatch problem of traditional retrieval methods~\cite{karpukhin2020dense}. 
After training, the models can be used to encode all documents in a target corpus for efficient ad hoc retrieval. 
Previous studies have shown that, when the target corpus is the same as or similar to the training corpus, i.e., the in-domain scenario, DR models can significantly outperform traditional first-stage retrieval methods like BM25 by a large margin~\cite{lin2021batch, karpukhin2020dense}.
DR models can not only deliver better performance on precision-oriented measures~\cite{xiong2021approximate, zhan2021optimizing, lin2021batch}, but also benefit second-stage reranking by recalling much more relevant documents~\cite{hofstatter2021efficiently, qu2021rocketqa, zhan2022learning}. 

However, recent studies find that the remarkable in-domain performance of DR does not translate into out-of-domain scenarios where the target domain is different from the domain of training data. 
DR is empirically vulnerable to domain shift and struggles to outperform traditional retrieval methods in out-of-domain scenarios~\cite{thakur2021beir, zhan2022evaluating, sciavolino2021simple}. 
Because out-of-domain inference is inevitable in situations where sufficient training data is not available in the target domain, this weakness would limit the employment of DR in real IR systems. 
For example, it is prohibitive to annotate relevant pairs in some medical domains due to privacy constraints or in some confidential situations due to security concerns. 
Hence it is important to make DR more adaptable to the domain shift. 

\begin{figure*}
	\subfloat[Dense Retrieval]{\label{fig:dr_conceptual}\includegraphics[height=0.14\linewidth]{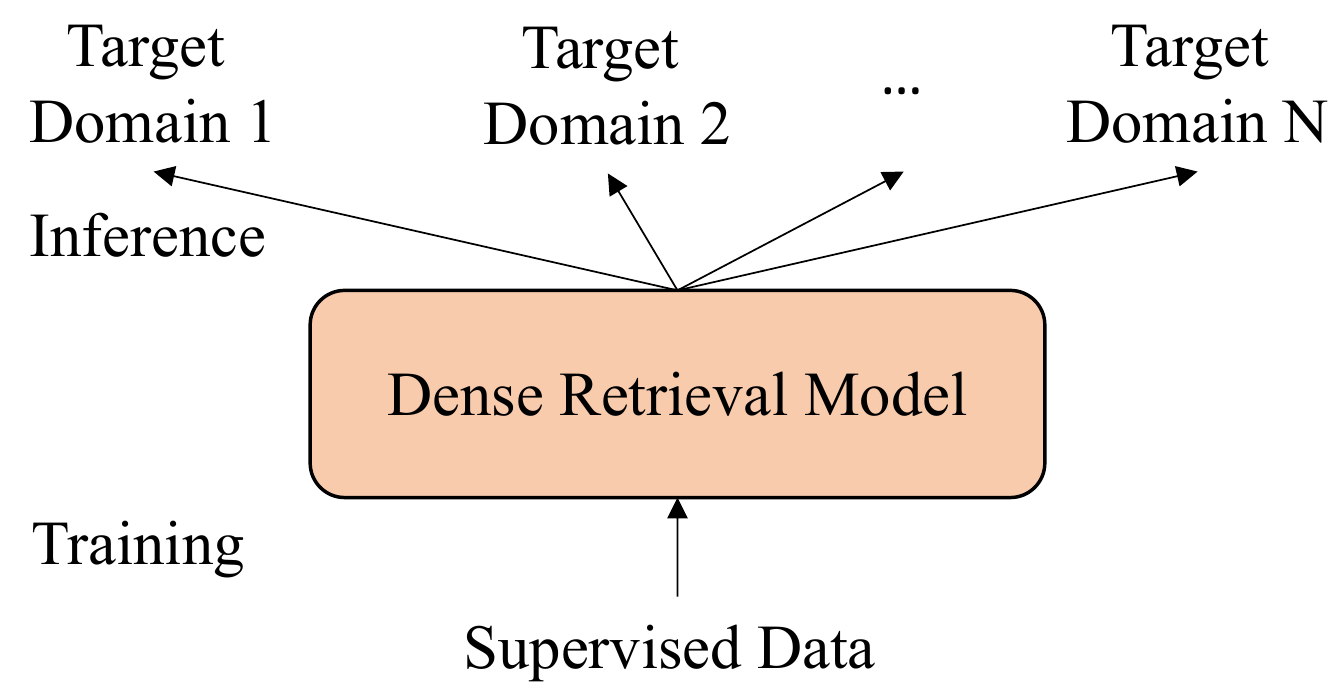}} \hfil  
	\subfloat[Disentangled Dense Retrieval]{\label{fig:ddr_conceptual}\includegraphics[height=0.14\linewidth]{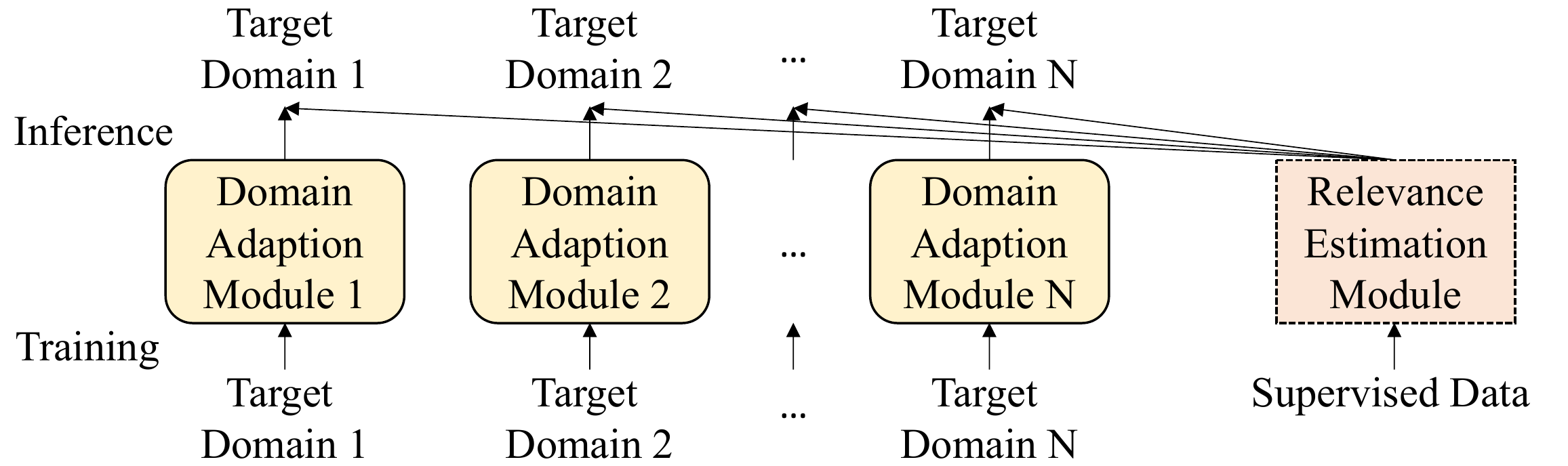}}  
    \caption{Schematic diagrams illustrating Dense Retrieval and the proposed Disentangled Dense Retrieval.\label{fig:compare_dr_ddr}}
\end{figure*}

We argue that the out-of-domain performance loss is to some extent caused by the differences in language features between the training and the target corpora. 
When finetuned with supervised data, DR learns not only how to estimate relevance between queries and documents, but also how to fit the features of the training domain such as word distribution, writing style, document structure, etc. These features might vary across domains. Thus the performance of DR degenerates in out-of-domain scenarios because it can not adapt to the language features of a new target domain. 
A possible solution, from this point of view, is a design that enables flexibly updating the domain modeling ability in different target domains while retaining the relevance estimation ability.
In fact, many classic retrieval models in the pre-dense-retrieval era have already followed this principle. Take BM25 as an example. It utilizes the same formula for estimating relevance scores across domains but measures word importance with corpus-specific IDF values~\cite{bm25}. 
However, the design does not exist in DR where the abilities of relevance estimation and domain modeling are jointly learned during training and entangled within the model parameters. 

In this paper, we propose a novel DR framework named Disentangled Dense Retrieval~(DDR), which disentangles relevance estimation and domain modeling to support effective and flexible domain adaption. Specifically, DDR consists of a Relevance Estimation Module~(REM) and one or more Domain Adaption Modules~(DAMs), both of which are optimized separately and are assembled during inference. REM is optimized with a labeled training corpus once and for all, and it does not require additional supervised data of the target domain. When applied to a new domain, DDR trains a separate DAM on the new corpus to mitigate domain shift.  

Figure~\ref{fig:compare_dr_ddr} contrasts the proposed DDR with the existing DR paradigm. 
On the left, Figure~\ref{fig:dr_conceptual} illustrates the training and evaluation process of DR. The model is trained once and is evaluated in various target domains. Since the model has never seen the target-domain corpus during training, ranking effectiveness often degenerates. For example, the model has trouble understanding the word `COVID-19' when it is not involved in the training data but appears in the target domain. 
On the right, Figure~\ref{fig:ddr_conceptual} shows the training and adaption process of DDR. REM is trained on supervised data and learns domain-invariant matching patterns. 
DAM is trained for each target domain to capture domain-specific features and thus helps DDR adapts to the target domain.
Compared with DR which attempts to apply the same model to various domains, DDR is more adaptable thanks to the flexible DAMs.  

In order to implement the DDR framework, we need to address the problems of how to disentangle domain modeling and relevance estimation in model architecture and how to appropriately train the two modules. In this paper, we employ the following implementation designs. 
Firstly, inspired by delta tuning~\cite{ding2022delta, hu2021lora, he2022towards}, we propose a disentangled architecture that uses the Transformer backbone as DAM and small insertable networks as REM. It enables disjoint parameters for the two abilities and induces marginal inference overhead. 
Secondly, for training REM, we propose a technique named disentangled finetuning. Through disentangled finetuning, we could prevent REM from fitting the domain-specific features, which helps it learn generic relevance matching patterns that could generalize to various target domains. 
Finally, for training DAM, we utilize a special initialization of its parameters to tackle the potential incompatibilities with REM caused by the disentanglement in architecture and training. After initialization, we use masked language modeling~\cite{devlin2019bert} as the optimization objective for training DAM in a target domain, which is unsupervised and does not require any domain-specific knowledge or annotations.

In experiments, we compare DDR with traditional methods and various DR baselines in different languages, in different domains, and with different finetuning methods. 
Experimental results show that DDR considerably boosts the out-of-domain ranking performance.
Not only does it outperform the traditional retrieval methods by a large margin in most scenarios, but it also substantially outperforms the DR counterpart under all settings and achieves 20\%-50\% relative improvement in some target domains.
We further conduct extensive ablation studies to validate the contributions of different designs.
\makeatletter
\if@ACM@anonymous
	The code and models will be publicly available to support reproducibility.
\else
	The code and models are available at \url{https://github.com/jingtaozhan/disentangled-retriever}
\fi
\makeatother

\section{Related Work}

DR represents queries and documents with embeddings and utilizes embedding similarities as estimated relevance scores. At the offline stage, it encodes all documents of a target corpus and builds an index of special infrastructure. During online serving, it encodes the user queries to embeddings and efficiently searches relevant candidates with approximate nearest neighbor search algorithms~\cite{zhan2022learning}. Transformer~\cite{vaswani2017attention} is usually used as the backbone. It is fully finetuned on labeled data to learn to encode queries and documents. Existing studies propose various finetuning methods, such as contrastive learning with hard negatives~\cite{xiong2021approximate, zhan2021optimizing} and knowledge distillation based on query-document relevance scores output by a strong cross-encoder~\cite{hofstatter2021efficiently, lin2021batch, qu2021rocketqa}. 
Recently, several studies challenge the generalization ability of DR. \citet{thakur2021beir} evaluate DR in several out-of-domain situations and find that its effectiveness substantially degenerates. \citet{zhan2022evaluating} resample the training data and observe that DR performs unsatisfactorily on queries that are dissimilar to the training queries. 
There are already several workarounds trying to tackle this problem. \citet{wang2022gpl} train the DR models based on pseudo labels in the target domain that are generated by a query generator and a cross encoder. \citet{xu2022laprador} and \citet{izacard2021unsupervised} pretrain DR models on crawled large-scale web pages with contrastive learning. 
Our proposed DDR is orthogonal to these methods. It does not require any pseudo labels or large-scale pretraining. By disentangling relevance estimation and domain modeling, DDR naturally facilitates effective and flexible domain adaption.
Besides, the existing workarounds can be integrated into DDR as more advanced optimization techniques for REM and DAM, which may yield further performance improvement.

\section{Disentangled Dense Retrieval}

In this section, we present Disentangled Dense Retrieval~(DDR), which supports unsupervisedly updating part of its parameters to effectively adapt to a specific domain. 
A conceptual diagram is illustrated in Figure~\ref{fig:ddr_conceptual}.
As shown in the figure, DDR employs two disentangled modules, i.e., a Relevance Estimation Module~(REM) and several Domain Adaption Modules~(DAMs). REM is trained on supervised data to learn generic matching patterns and is able to generalize to various target domains. DAM is updated on the target-domain corpus to mitigate the performance loss caused by domain shift. Thanks to them, DDR is capable of effectively adapting to an unseen domain. 
In the following, we will elaborate on the implementation designs, i.e., the disentangled model architecture and the techniques for optimizing the two modules.

\subsection{Disentangled Architecture}

\begin{figure}
    \includegraphics[width=0.85\linewidth, keepaspectratio=True]{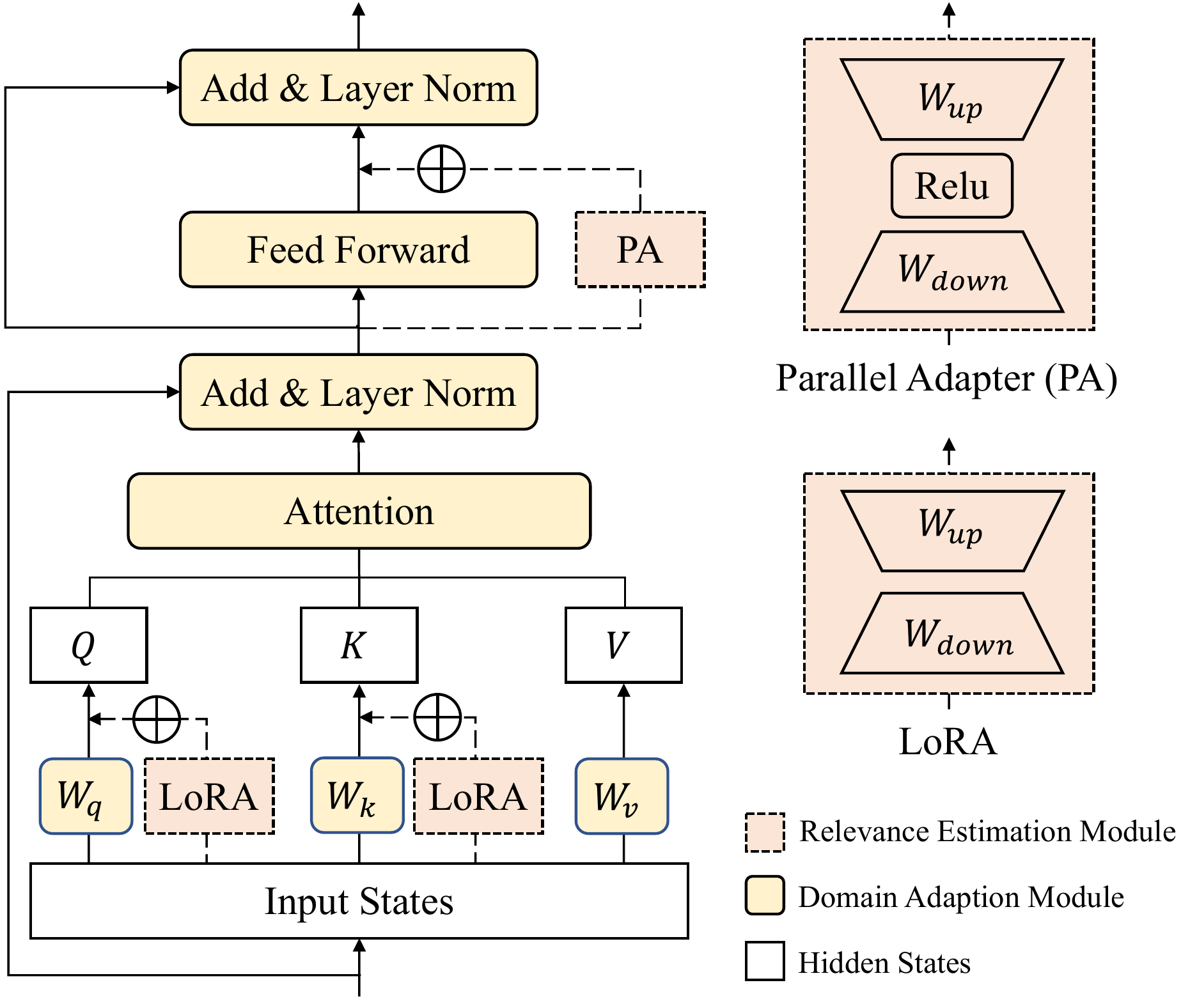}
    \caption{The disentangled architecture of DDR. 
    } 
    \label{fig:ddr_disentangle_architecture}
\end{figure}

An effective DDR architecture should not only disentangle relevance estimation and domain modeling but also possesses strong modeling capacity and marginal inference overhead. To achieve this, we carefully design a disentangled architecture, which is depicted in Figure~\ref{fig:ddr_disentangle_architecture}. 

As Figure~\ref{fig:ddr_disentangle_architecture} shows, the Transformer backbone is used for DAM.
Transformer employs a multi-layer architecture. 
Each layer employs a self-attention network, a feed forward network, and residual connections.
With suitable self-supervised objectives, the Transformer trained on a large corpus exhibit impressive capacities in fitting language features such as syntactic relations~\cite{clark-etal-2019-bert} and language structures\cite{jawahar2019does}. 
Since domain modeling shares a similar goal, i.e., fitting domain-specific features like word distribution, the Transformer backbone is a favorable option for implementing DAM. 

As for REM, inspired by delta tuning methods~\cite{jung2022semi, hu2021lora, he2022towards}, we utilize several small and insertable neural networks. As shown in Figure~\ref{fig:ddr_disentangle_architecture}, REM consists of LoRA~\cite{hu2021lora} and Parallel Adapter~(PA)~\cite{he2022towards}, both of which process hidden states with a bottleneck structure and add the output back to the backbone. They have been demonstrated to be effective at learning task information~\cite{hu2021lora, he2022towards}: only training the two networks yield performance comparable to fully finetuning the entire Transformer in some NLP tasks. 
Besides the effectiveness, they are also inference-efficient. LoRA can be directly merged into attention weights and causes zero inference overhead. PA consists of tiny parameters and the overhead is marginal. 
Given these advantages, they are a suitable choice for implementing REM.

\begin{figure}
    \includegraphics[width=1.0\linewidth, keepaspectratio=True]{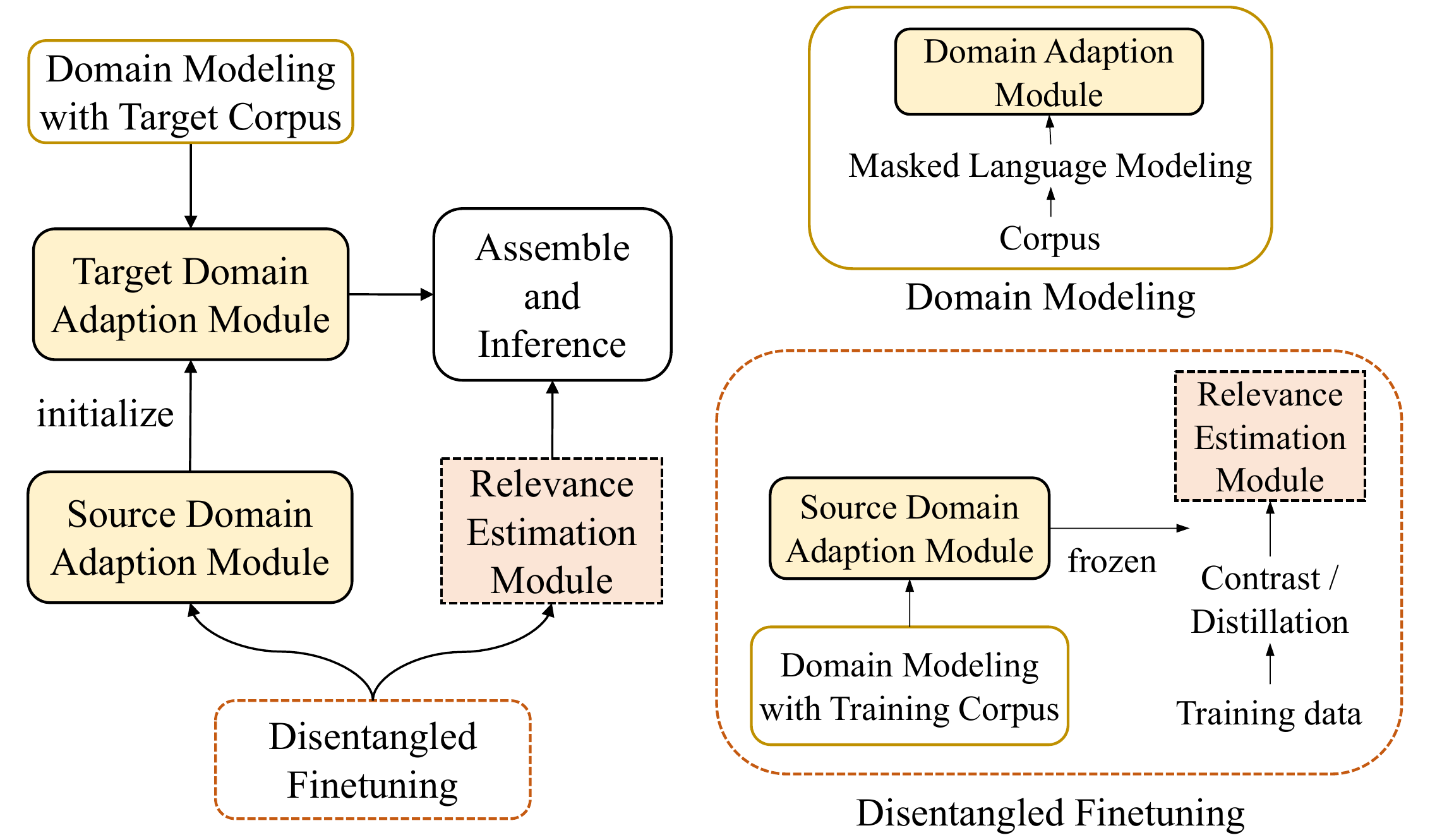}
    \caption{Training and adaption procedure of DDR. 
    } 
    \label{fig:ddr_train_procedure}
\end{figure}

\subsection{Optimizing Relevance Estimation Module}

Now we present how to train REM to learn relevance estimation. It is trained only once on supervised data and is directly used in unseen domains. We propose a simple yet effective \emph{disentangled finetuning} technique to improve its generalization ability in various domains. The technique aims to make REM focus on generic relevance matching patterns~(e.g., exact match and soft match) and overlook the domain-specific features~(e.g., document structure). It employs a two-step training process and is illustrated in Figure~\ref{fig:ddr_train_procedure}. 
Firstly, it trains a DAM to fit the training corpus. Secondly, it freezes the trained DAM and inserts the randomly initialized REM. REM is then trained with supervised data using DR-related loss functions, e.g., a contrastive loss or a knowledge distillation loss.
In the first step, DAM fits the training corpus via unsupervised learning and the domain-specific features are captured.
Thus, in the second step, REM can focus on complementary knowledge, i.e., relevance matching patterns, via supervised learning. It does not need to fit those domain features which have already been captured and transformed into layers of hidden semantic representations by DAM. Hence REM can still work well in a domain that is different from the domain of labeled data.

\subsection{Optimizing Domain Adaption Module}

Intuitively, it is challenging to train DAM, which should not only effectively model the target-domain features without supervision, but also be compatible with REM that is separately trained. To tackle the two challenges, we utilize a strong self-supervised objective and adopt a special initialization scheme.

We use masked language modeling~\cite{devlin2019bert} as the training objective, which has been demonstrated to help model effectively learn language features~\cite{clark-etal-2019-bert, jawahar2019does}. It masks part of the document tokens at random, and DAM is required to reconstruct the masked text. It needs neither supervised data nor domain-specific knowledge, which eases the adaption process. In our implementation, we stick to the parameter setting of BERT~\cite{devlin2019bert}. Specifically, $15\%$ document tokens are randomly chosen, of which $80\%$ are replaced by the special token `[MASK]', $10\%$ are replaced with a random token, and the rest $10\%$ are kept unchanged.
It is worth noting that there could be other objective options. For example, we can extract text spans and utilize contrastive pretraining~\cite{gao2021unsupervised, izacard2021unsupervised}, which is shown to be particularly effective for DR. We can also introduce IR-specific knowledge by training with pseudo queries~\cite{ma2021prop}, pseudo labels~\cite{wang2022gpl}, or IR axioms\cite{chen2022axiomatically}. 
Compared with these, masked language modeling requires much less human effort or knowledge when building training data and hence is adopted in this paper. We leave exploring other objectives to future studies. 

Although we can unsupervisedly train DAM in the target domain, a downside is that we create a gap between REM and DAM because they are separately trained and may not work well when assembled. To mitigate this, we adopt a simple but effective technique named \emph{sequential initialization}. Recall that REM is trained alongside a frozen DAM in the previous section. In Figure~\ref{fig:ddr_train_procedure}, this frozen DAM is called the source DAM. The source DAM is compatible with REM because they see each other during training. Therefore, we use the source DAM as an initialization point and a relatively small learning rate when training the target DAM in the target domain. In this way, the target DAM is similar to the source DAM and its compatibility with REM improves. Despite the simplicity, sequential initialization works impressively well, yielding robust performance improvement in our experiments.

\section{Experimental Setup}

In this section, we introduce our experimental settings. We select out-of-domain test sets across multiple domains and in both English and Chinese. A variety of baselines are implemented or adopted, including traditional methods, self-implemented DR models that follow the same finetuning process as DDR, and the strong DR models in the literature. We compare DR model performance when finetuned by either contrastive learning or knowledge distillation. We will introduce the setup details in the following sections.

\begin{table}[t]
	\small
    \centering
    \caption{Statistics about datasets used to evaluate out-of-domain retrieval performance. The table presents the number of documents/queries and their average lengths. `en' and `zh' are abbreviations for English and Chinese, respectively. }
    \label{tab:dataset_statistics}
    \begin{tabular}{ll|rrrr}
    \toprule
        \multicolumn{2}{l|}{\multirow{2}{*}{Dataset}} & \multicolumn{2}{c}{Document} & \multicolumn{2}{c}{Query} \\ 
        \multicolumn{2}{l|}{} & \multicolumn{1}{c}{\#Num} & \multicolumn{1}{c}{\#Word} & \multicolumn{1}{c}{\#Num} & \multicolumn{1}{c}{\#Word} \\ \midrule
        \multirow{6}{*}{en}
        & TREC-Covid 		& 171,332 	& 161 & 50    & 11\\ 
        & Lotte-Write 		& 199,994 	& 120 & 2,000 & 10 \\ 
        & Lotte-Rec 		& 166,975 	& 138 & 2,002 & 10 \\ 
        & Lotte-Tech 		& 638,509 	& 145 & 2,004 & 10 \\ 
        & Lotte-Life 		& 119,461 	& 153 & 2,002 & 11 \\ 
        & Lotte-Sci			& 1,694,164 & 128 & 2,017 & 9  \\ 
        \midrule
        \multirow{4}{*}{zh} 
        & CPR-Ecom 			& 1,002,822 & 26 & 1,000  & 6 \\ 
       	& CPR-Video			& 1,000,000 & 20 & 1,000  & 7 \\ 
        & CPR-Medical 		& 960,526 	& 107 & 1,000 & 17 \\ 
        & cMedQAv2			& 226,266 	& 90 & 4,000  & 43\\ 
    \bottomrule
    \end{tabular}
\end{table}

\subsection{Datasets}

We first present the training datasets. We use MS MARCO passage ranking dataset~\cite{Craswell2021MSMB} as English training data and Dureader~\cite{he2017dureader} as Chinese training data. Both datasets contain passages from web pages and queries from search logs. The former contains 0.5 million training queries and 8.8 million passages. 
The latter is originally proposed as a Question-Answering dataset. It is transformed into a retrieval dataset by treating the passage labeled `most related to the query' as the positive passage. 
The transformed retrieval dataset contains 0.3 million training queries and 8.5 million passages. 

Next, we introduce the out-of-domain test sets.
For English experiments, we use Lotte benchmark~\cite{santhanam2021colbertv2} and TREC-Covid dataset~\cite{voorhees2021trec}. Lotte collects questions and answers posted on StackExchange and divides them into five topics including writing, recreation, science, technology, and lifestyle. It regards the accepted or upvoted answers as relevant. 
TREC-Covid is a retrieval dataset about searching COVID-19-related information from biomedical literature articles. The original annotation is biased towards lexical retrieval methods and we utilize a more comprehensive annotation released by \citet{thakur2021beir}.
For Chinese experiments, we utilize CPR benchmark~\cite{long2022multi} and cMedQAv2~\cite{zhang2018multi}. CPR benchmark consists of three human-annotated domain-specific retrieval datasets collected from an e-commerce platform~(Taobao), a video platform~(Youku), and medical search within a search engine~(Quark). cMedQAv2 is constructed based on an online Chinese medical question answering forum~(\url{http://www.xywy.com}). It collects user descriptions of their symptoms and the diagnosis or suggestions responded by doctors. We regard user descriptions as queries and responses as relevant documents. 
The dataset statistics are summarized in Table~\ref{tab:dataset_statistics}. 

\subsection{Baselines}

We provide traditional retrieval baselines and DR baselines. The former consist of BM25~\cite{bm25} and BM25+RM3~\cite{abdul2004umass}, which are implemented based on Anserini~\cite{yang2018anserini}. The latter consists of self-implemented DR models and the competitive DR models in the literature. 

\begin{table*}[t]
	\small
    \centering
    \caption{Full retrieval performance on English and Chinese retrieval datasets. The DR baselines share the same backbone, pooling strategy, and finetuning settings as DDR.
    */** indicates statistically significant improvement at $p < 0.05/0.01$ level measured by two-tailed pairwise t-test. The best results are marked in bold. The second-best results are underlined.}
    \label{tab:main_comparison}
    \begin{tabular}{l|ll|C{10mm}C{10mm}|C{10mm}C{10mm}R{10mm}|C{10mm}C{10mm}R{10mm}}
    \toprule
        \multirow{2}{*}{Language} & \multirow{2}{*}{Datasets} & \multirow{2}{*}{Metrics} & \multirow{2}{*}{BM25} & \multirow{2}{*}{+RM3} & \multicolumn{3}{c|}{Contrastive Learning} & \multicolumn{3}{c}{Knowledge Distillation} \\ 
         &  &  & &  & DR & DDR & \multicolumn{1}{c|}{imp.} & DR & DDR & \multicolumn{1}{c}{imp.} \\ \hline
        \multirow{18}{*}{English} 
         & \multirow{3}{*}{TREC-Covid} 
         & NDCG@10 	& 0.603 & 0.598 & 0.648  & \textbf{0.720}  & 11.1\%\,\,\;\; & 0.679  & \underline{0.714}  & 5.2\%\,\,\;\; \\ 
         & & R@100 	& 0.109 & 0.115 & 0.093  & \underline{0.116}  & 24.7\%\,** & 0.108  & \textbf{0.122}  & 13.0\%\,** \\ 
         & & R@1000 & 0.395 & \textbf{0.443} & 0.315  & 0.432  & 37.1\%\,** & 0.355  & \underline{0.434}  & 22.3\%\,** \\ \cline{2-11}
         & \multirow{3}{*}{Lotte-Write} 
         & NDCG@10 	& 0.352 & 0.338 & 0.352  & \textbf{0.447}  & 27.0\%\,** & 0.392  & \underline{0.435}  & 11.0\%\,** \\ 
         & & R@100 	& 0.541 & 0.557 & 0.550  & \textbf{0.663}  & 20.5\%\,** & 0.599  & \underline{0.658}  & 9.8\%\,** \\ 
         & & R@1000 & 0.681 & 0.703 & 0.707  & \textbf{0.805}  & 13.9\%\,** & 0.734  & \underline{0.791}  & 7.8\%\,** \\ \cline{2-11}
         & \multirow{3}{*}{Lotte-Rec} 
         & NDCG@10 	& 0.325 & 0.294 & 0.360  & \underline{0.411}  & 14.2\%\,** & 0.398  & \textbf{0.424}  & 6.5\%\,** \\ 
         & & R@100 	& 0.570 & 0.590 & 0.592  & \textbf{0.701}  & 18.4\%\,** & 0.651  & \textbf{0.701}  & 7.7\%\,** \\ 
         & & R@1000 & 0.736 & 0.772 & 0.769  & \textbf{0.866}  & 12.6\%\,** & 0.797  & \underline{0.849}  & 6.5\%\,** \\ \cline{2-11}
         & \multirow{3}{*}{Lotte-Tech} 
         & NDCG@10 	& 0.151 & 0.146 & 0.141  & \textbf{0.207}  & 46.8\%\,** & 0.175  & \underline{0.204}  & 16.6\%\,** \\ 
         & & R@100 	& 0.335 & 0.334 & 0.308  & \textbf{0.450}  & 46.1\%\,** & 0.364  & \underline{0.438}  & 20.3\%\,** \\ 
         & & R@1000 & 0.539 & 0.575 & 0.529  & \textbf{0.711}  & 34.4\%\,** & 0.586  & \underline{0.698}  & 19.1\%\,** \\ \cline{2-11}
         & \multirow{3}{*}{Lotte-Life} 
         & NDCG@10 	& 0.305 & 0.270 & 0.375  & \underline{0.431}  & 14.9\%\,** & 0.409  & \textbf{0.433}  & 5.9\%\,** \\ 
         & & R@100 	& 0.548 & 0.544 & 0.651  & \textbf{0.719}  & 10.4\%\,** & 0.682  & \underline{0.717}  & 5.1\%\,** \\ 
         & & R@1000 & 0.754 & 0.776 & 0.846  & \textbf{0.903}  & 6.7\%\,** & 0.866  & \underline{0.899}  & 3.8\%\,** \\ \cline{2-11}
         & \multirow{3}{*}{Lotte-Sci} 
         & NDCG@10 	& \textbf{0.156} & 0.144 & 0.121  & {0.149}  & 23.1\%\,** & 0.143  & \underline{0.151}  & 5.6\%\,*\,\, \\ 
         & & R@100 	& 0.303 & 0.293 & 0.237  & \underline{0.308}  & 30.0\%\,** & 0.283  & \textbf{0.316}  & 11.7\%\,** \\ 
         & & R@1000 & 0.498 & 0.504 & 0.398  & \underline{0.520}  & 30.7\%\,** & 0.468  & \textbf{0.531}  & 13.5\%\,** \\ \hline
        \multirow{12}{*}{Chinese} 
         & \multirow{3}{*}{CPR-Ecom} 
         & NDCG@10 	& 0.255 & 0.222 & 0.263  & \underline{0.313}  & 19.0\%\,** & 0.287  & \textbf{0.318}  & 10.8\%\,** \\ 
         & & R@100 	& 0.582 & 0.616 & 0.623  & \textbf{0.703}  & 12.8\%\,** & 0.647  & \underline{0.682}  & 5.4\%\,** \\ 
         & & R@1000 & 0.807 & 0.841 & 0.843  & \textbf{0.882}  & 4.6\%\,** & 0.824  & \underline{0.861}  & 4.5\%\,** \\ \cline{2-11}
         & \multirow{3}{*}{CPR-Video} 
         & NDCG@10 	& \underline{0.284} & 0.259 & 0.251 & 0.269 & 7.2\%\,** & 0.284  & \textbf{0.298}  & 4.9\%\,*\,\, \\ 
         & & R@100 	& \textbf{0.738} & 0.724 & 0.639  & 0.684  & 7.0\%\,** & 0.706  & \underline{0.737}  & 4.4\%\,** \\ 
         & & R@1000 & \textbf{0.900} & 0.885 & 0.825  & 0.859  & 4.1\%\,** & 0.858  & \underline{0.887}  & 3.4\%\,** \\ \cline{2-11}
         & \multirow{3}{*}{CPR-Medical} 
         & NDCG@10 	& 0.238 & 0.207 & 0.303  & \textbf{0.324}  & 6.9\%\,** & 0.285  & \underline{0.302}  & 6.0\%\,** \\ 
         & & R@100 	& 0.393 & 0.350 & 0.491  & \textbf{0.502}  & 2.2\%\,\,\;\; & 0.476  & \underline{0.488}  & 2.5\%\,\,\;\; \\ 
         & & R@1000 & 0.510 & 0.488 & 0.636  & \textbf{0.668}  & 5.0\%\,** & 0.612  & \underline{0.643}  & 5.1\%\,** \\ \cline{2-11}
         & \multirow{3}{*}{cMedQAv2} 
         & NDCG@10 	& 0.099 & 0.076 & \underline{0.110}  & \textbf{0.114}  & 3.6\%\,\,\;\; & 0.107  & 0.108  & 0.9\%\,\,\;\; \\ 
         & & R@100 	& 0.210 & 0.190 & \underline{0.262}  & \textbf{0.282}  & 7.6\%\,** & 0.249  & \underline{0.262}  & 5.2\%\,** \\ 
         & & R@1000 & 0.390 & 0.394 & 0.499  & \textbf{0.532}  & 6.6\%\,** & 0.468  & \underline{0.500}  & 6.8\%\,** \\ 
    \bottomrule
    \end{tabular}
\end{table*}

We implement and train DR models that share the same backbone, pooling strategy, and finetuning parameter settings as DDR. Therefore, we can rule out the potential influence caused by different implementation details. Concretely, the self-implemented DR baselines are initialized by BERT-base and utilize average pooling when outputting the sequence embedding. They share the same finetuning settings such as batch size, training step, negative sampling, distillation labels, etc, except for the learning rate which is tuned and set to $1 \times 10 ^{-5}$.
They are finetuned on MS MARCO or Dureader with contrastive learning or distillation and are then evaluated in target domains in a zero-shot manner. 
The comparison between DDR and the self-implemented DR models is the main focus of our experiments. 

We also report the performance of strong DR models in the literature to put our results into context. They include ANCE~\cite{xiong2021approximate}, ADORE~\cite{zhan2021optimizing}, TAS-B\cite{hofstatter2021efficiently}, RocketQA~\cite{qu2021rocketqa}, TCT v2~\cite{lin2021batch}, coCon~\cite{gao2021unsupervised}, and Contriever~\cite{izacard2021unsupervised}. They usually adopt more complex training settings such as multi-turn hard negative mining~\cite{xiong2021approximate, zhan2021optimizing}, ensembled distillation labels~\cite{hofstatter2021efficiently}, contrastive pretraining~\cite{izacard2021unsupervised, gao2021unsupervised}. 
Note that these training techniques are orthogonal to our proposed disentangled modeling and can be employed to optimize DAM or REM for further performance improvement in the future.

\subsection{Implementation Details}

Implementation is largely consistent across languages and domains unless otherwise stated.

We average the output token embeddings as the sequence embedding. We use inner-product to compute embedding similarities in the English experiments following the common practice on MS MARCO dataset~\cite{xiong2021approximate, izacard2021unsupervised} but cosine similarity in the Chinese experiments because it performs better.  
The backbone is initialized from bert-base-uncased/chinese checkpoints~\cite{devlin2019bert}.

When we train DAM, the batch size and max input length are $1,024$ and $100$ for datasets with short documents and $512$ and $190$ for datasets with long documents, respectively. 
The learning rate is $5 \times 10^{-5}$ and $2 \times 10^{-5}$ during disentangled finetuning and target-domain adaption, respectively.

When contrastive learning is used for finetuning REM, a mini-batch consists of $128$ queries and the corresponding $128$ positive documents. On MS MARCO, we additionally add $3 \times 128$ hard negatives to the mini-batch. On Chinese Dureader, We only use in-batch negatives because the dataset has too many false negatives. The hard negatives on MS MARCO are downloaded from the public resource\footnote{\label{note:sbert}\url{https://huggingface.co/datasets/sentence-transformers/msmarco-hard-negatives}}. They are irrelevant top ranking results of multiple retrievers. The learning rate is $2 \times 10^{-5}$. Model performance converges after $5$ epochs on MS MARCO and $1$ epoch on Dureader.

\newcommand{\sigc}{*}
\newcommand{\sigd}{$^\circ$} 
\newcommand{\sigcd}{\sigc\sigd}
\newcommand{\emptysigcd}{\;\;\;}

\begin{table*}[t]
	\small
    \centering
    \caption{Comparing DDR with competitive DR models in the literature. We report R@1000 on different domains. Best results are marked bold. Second-best results are underlined. \sigc and \sigd indicate that DDR~(contrast) and DDR~(distill) significantly outperform the baseline, respectively (paired t-test, $p < 0.01$). `en' is the abbreviation for English.}
    \label{tab:compare_with_dr_literature}
    \begin{tabular}{l|C{10mm}C{10mm}cC{10mm}C{10mm}C{10mm}c|C{17mm}C{14mm}}
    \toprule
        Models & ANCE & ADORE & RocketQA & TAS-B & TCT v2 & coCon & Contriever & DDR~(contrast) & DDR~(distill) \\ 
        Contrastive Pretraining &  &  &  &  &  & $\checkmark$ & $\checkmark$ &  & \\ 
        Additional Resources & & & search log & & & & web pages & & \\ \midrule
        TREC-Covid~(en) 		& 0.348\sigcd  & 0.346\sigcd  & 0.364\sigcd  & 0.333\sigcd  & 0.353\sigcd  & 0.400\,\,\sigd  & 0.344\sigcd  & \underline{0.432}  & \textbf{0.434}  \\
         
        Lotte-Write~(en) 		& 0.700\sigcd  & 0.736\sigcd  & 0.547\sigcd  & 0.712\sigcd  & 0.726\sigcd  & 0.744\sigcd  & 0.777\sigcd  & \textbf{0.805}  & \underline{0.791}  \\ 
        
        Lotte-Rec~(en) 			& 0.775\sigcd  & 0.790\sigcd  & 0.495\sigcd  & 0.811\sigcd  & 0.797\sigcd  & 0.823\sigcd  & \underline{0.857}\emptysigcd  & \textbf{0.866}  & 0.849  \\ 
        
        Lotte-Tech~(en) 		& 0.574\sigcd  & 0.594\sigcd  & 0.639\sigcd  & 0.595\sigcd  & 0.577\sigcd  & 0.611\sigcd  & 0.662\sigcd & \textbf{0.711}  & \underline{0.698}  \\ 
        
        Lotte-Life~(en) 		& 0.854\sigcd  & 0.868\sigcd  & 0.892\sigc\;\,  & 0.873\sigcd  & 0.870\sigcd  & 0.856\sigcd  & \textbf{0.909}\emptysigcd  & \underline{0.903}  & 0.899  \\ 
        
        Lotte-Sci~(en) 			& 0.451\sigcd  & 0.462\sigcd  & 0.504\sigcd  & 0.464\sigcd  & 0.446\sigcd  & 0.374\sigcd  & 0.498\sigcd  & \underline{0.520}  & \textbf{0.531}  \\ 
    \bottomrule
    \end{tabular}
\end{table*}

\let\sigc\undefined
\let\sigd\undefined
\let\sigcd\undefined
\let\emptysig\undefined
\let\emptysigcd\undefined

When knowledge distillation is used for finetuning REM, a mini-batch consists of $128$ queries, the corresponding $128$ positive documents, and $3 \times 128$ negatives. The loss function is Margin-MSE\cite{hofstatter2021efficiently}. The learning rate is $2 \times 10^{-5}$. On MS MARCO, the distillation labels are downloaded from the public resource\footnoteref{note:sbert}. They are the output scores of a cross-encoder. 
On Chinese Dureader, since no distillation labels are available, we train a cross-encoder and then use it to score the query-passage pairs. Model performance converges after $20$ epochs on MS MARCO and $1$ epoch on Dureader.

\section{Experimental results}

This section presents the empirical results. We first study the advantage of DDR by comparing it with DR. Then we conduct ablation studies to investigate the contribution of various components.

\subsection{Comparison between DR and DDR}

In this section, we comprehensively evaluate DDR. We employ both self-trained DR models and the strong DR models in the literature. The former follows exactly the same finetuning procedure as DDR, which enables fair comparison because finetuning is a sensitive factor of DR's effectiveness. The latter helps to put our results into context and further highlights the advantage of DDR even compared with sophisticated DR models. 

We first present our main experimental results in Table~\ref{tab:main_comparison}, which shows the full retrieval performance of traditional methods, self-trained DR models, and the proposed DDR models. We can see that DDR substantially boosts the ranking performance by a large margin. Specifically, we have the following three observations.
\begin{itemize}[leftmargin=*, topsep=0pt] %
	\item Firstly, DDR substantially outperforms traditional methods in most out-of-domain scenarios, while DR struggles. According to R@1000 results, DR underperforms BM25 on TREC-Covid, Lotte-Tech, Lotte-Sci, and CPR-Video when finetuned with contrastive loss. With a more complex distillation process, DR is able to achieve better performance but still underperforms BM25 on TREC-Covid and CPR-Video. Such phenomenon is consistent with the findings of \citet{thakur2021beir}. Nevertheless, the issue is largely resolved with the help of DDR. After finetuned with a contrastive loss, DDR is capable of outperforming BM25 by a large margin on most datasets except for CPR-Video. With distillation finetuning, the performance of DDR on CPR-Video further improves and is comparable to that of BM25. 
	\item Secondly, DDR considerably outperforms DR in terms of precision- and recall-oriented measures. The improvement is substantial and statistically significant. It is also robust to different languages, domains, and finetuning methods. For instance, on the Lotte-Tech test set, DDR can improve NDCG and Recall by up to $40\%$ when finetuned by a contrastive loss and $20\%$ when finetuned with knowledge distillation. Therefore, DDR effectively addresses the vulnerability of DR in out-of-domain scenarios.
	\item Thirdly, knowledge distillation seems to be particularly effective for DR but not for DDR. DR finetuned with knowledge distillation is generally more effective than contrastive learning, whereas DDR exhibits similar ranking effectiveness in both situations. 
	We speculate that differences in model architectures might result in inconsistent behavior and leave this to future studies. 
\end{itemize}

Next, we compare DDR with competitive DR models in the literature. We evaluate these models using their public checkpoints and report R@1000 results in Table~\ref{tab:compare_with_dr_literature}. Note that we do not report performance on Chinese datasets because the baselines are all developed in English. Now we discuss the experimental results by comparing DDR with different categories of baselines. 
\begin{itemize}[leftmargin=*, topsep=0pt] %
	\item Firstly, DDR considerably outperforms DR models that employ sophisticated hard negative mining strategies, such as ANCE~\cite{xiong2021approximate} and ADORE~\cite{zhan2021optimizing}. ANCE asynchronously updates the hard negative candidates during training and ADORE synchronously retrieves the hardest negatives at each training step. Compared to them, our contrastive finetuning process simply utilizes a static set of hard negatives, but this simple process still yields remarkable out-of-domain performance gains thanks to the strong domain adaption ability of DDR. We also believe that the sophisticated sampling methods of ANCE and ADORE can further advance the effectiveness of DDR and will explore it in the future.
	\item Secondly, DDR substantially outperforms DR models that are finetuned with knowledge distillation, such as RocketQA~\cite{qu2021rocketqa}, TAS-B~\cite{hofstatter2021efficiently}, and TCT v2~\cite{lin2021batch}. RocketQA additionally trains the DR model on a search log, while TAS-B and TCT v2 employ a distillation process like ours. Although the three baselines outperform ANCE and ADORE, they are significantly less effective than DDR. These models never see the target domains during training, while DDR is capable of updating DAM based on the target-domain corpus to actively adapt to the new domain. As a result, DDR delivers significant performance improvement.
	\item Thirdly, DDR achieves significantly better or similar effectiveness compared with DR models that are contrastively pretrained, such as coCon~\cite{gao2021unsupervised} and Contriever~\cite{izacard2021unsupervised}. coCon is pretrained on the training corpus and Contriever is pretrained on crawled large-scale web pages. Results show that contrastive pretraining helps them outperform other DR baselines. However, pretraining techniques employed by coCon and Contriever can hardly mitigate domain shift for the target domain that is not involved in the pretraining corpus. As shown in the results, Contriever performs unstably across domains: it is comparable with DDR on Lotte-Rec and Lotte-Life but is significantly less effective on other datasets, which we believe is due to different similarities between the pretraining corpus and the target-domain corpus. On the contrary, DDR yields stable performance improvement because it directly trains DAM on the target-domain corpus. Therefore, DDR is a better choice for first-stage retrieval. Besides, it is worth noting that masked language modeling utilized by DDR is demonstrated to be less effective for DR models than contrastive pretraining~\cite{gao2021unsupervised, izacard2021unsupervised}. Thus DDR may be further improved if contrastive learning is the optimization objective of DAM. We leave this to future studies. 
\end{itemize}

\subsection{Ablation Study}

Now we conduct extensive ablation studies to deeply analyze DDR. 
We will first show the benefits of the DDR framework, i.e., how updating DAM improves the out-of-domain effectiveness and how disentangled modeling facilitates efficient domain adaption. 
Then we will dive into the implementation designs and disclose the impact of model architectures and optimization techniques.

\subsubsection{Contribution of Adapting DAM} \mbox{}

\begin{figure}[t]
	\subfloat{\includegraphics[width=0.33\linewidth]{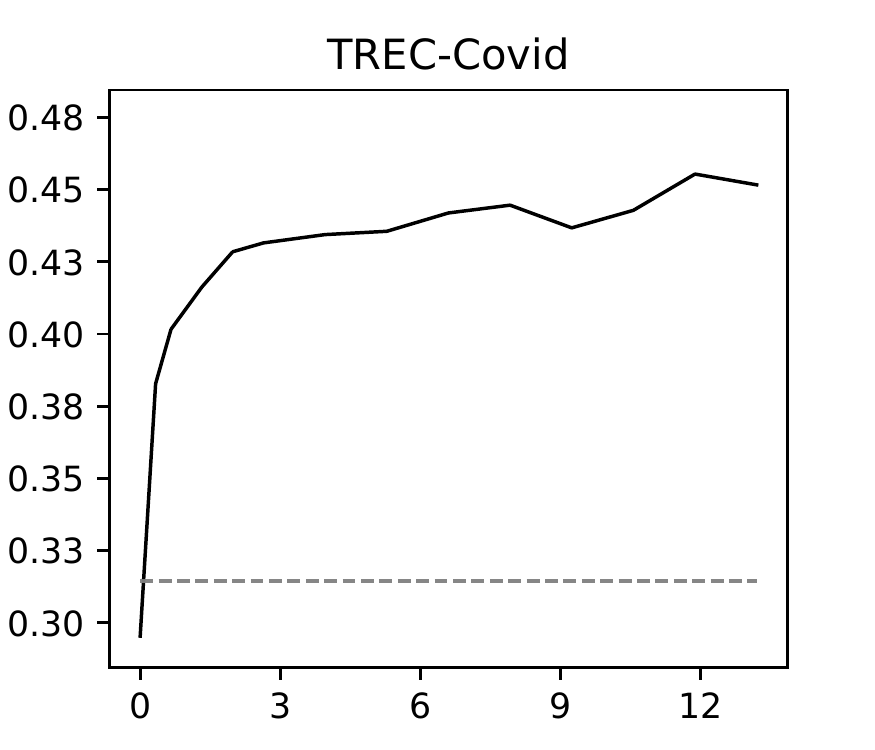}}   
	\subfloat{\includegraphics[width=0.33\linewidth]{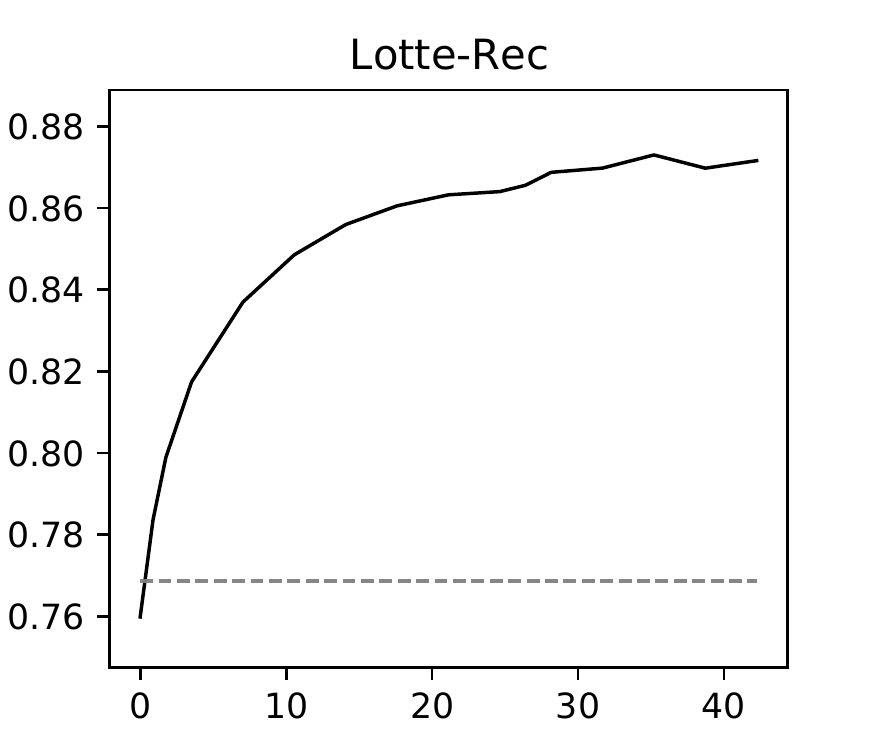}}  
	\subfloat{\includegraphics[width=0.33\linewidth]{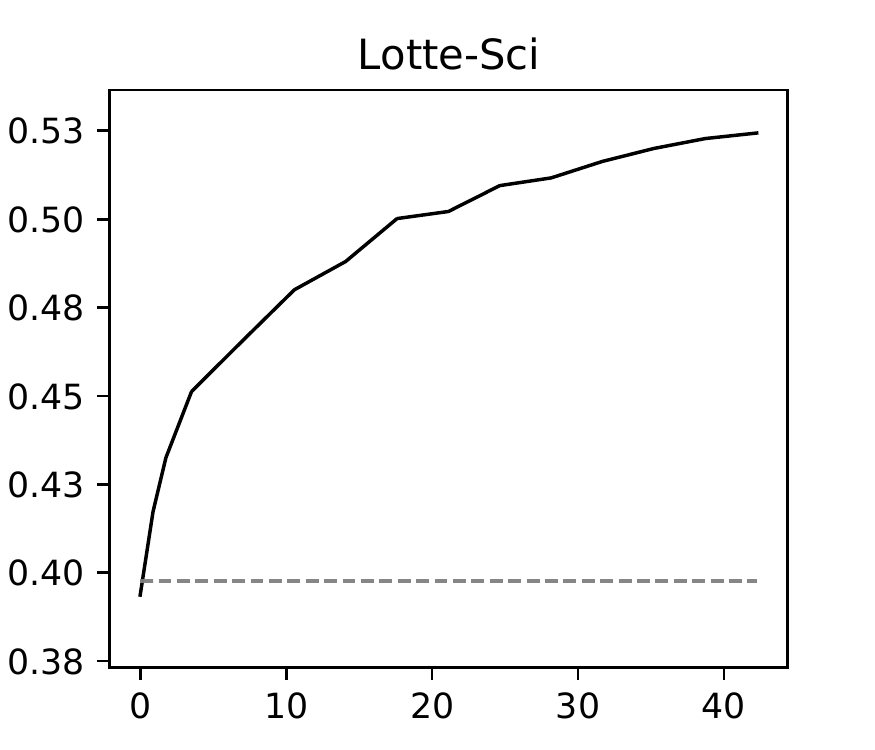}} 
	\\  
	\subfloat{\includegraphics[width=0.33\linewidth]{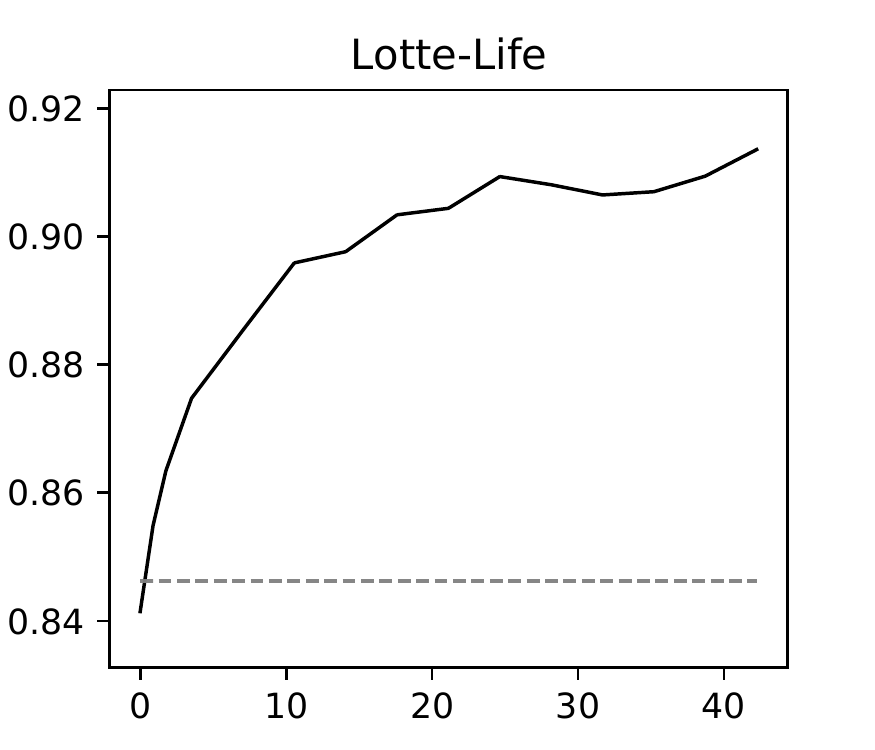}} 
	\subfloat{\includegraphics[width=0.33\linewidth]{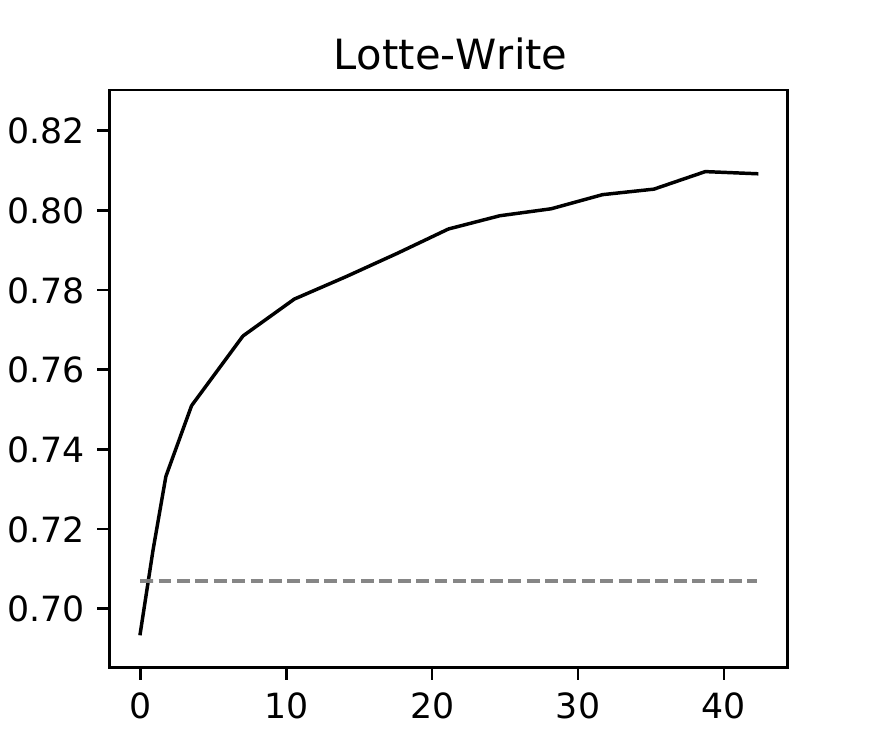}} 
	\subfloat{\includegraphics[width=0.33\linewidth]{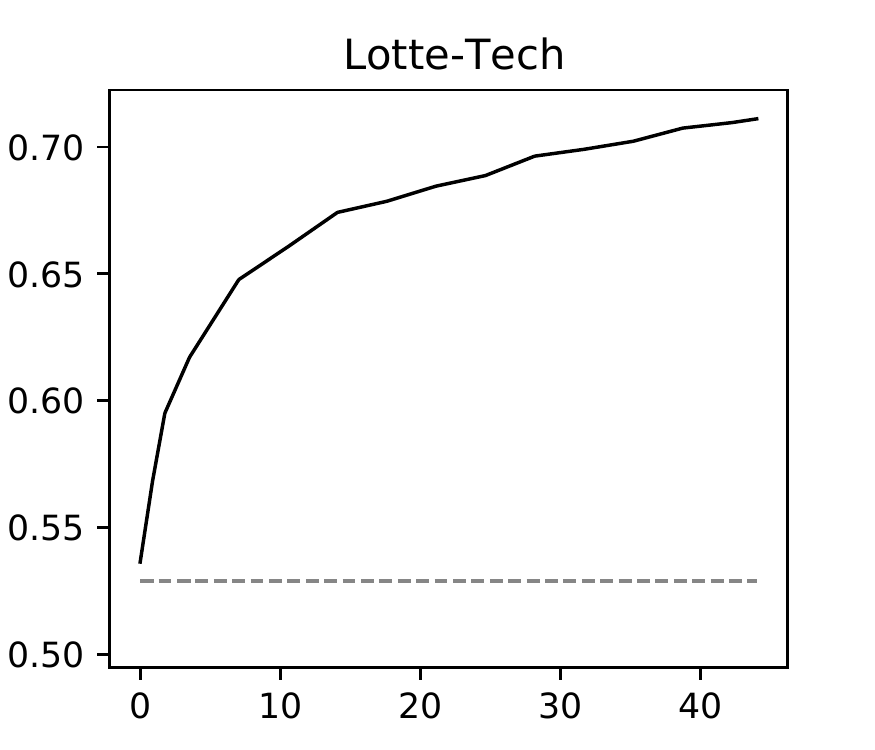}} 
    \\
	\subfloat{\includegraphics[width=0.33\linewidth]{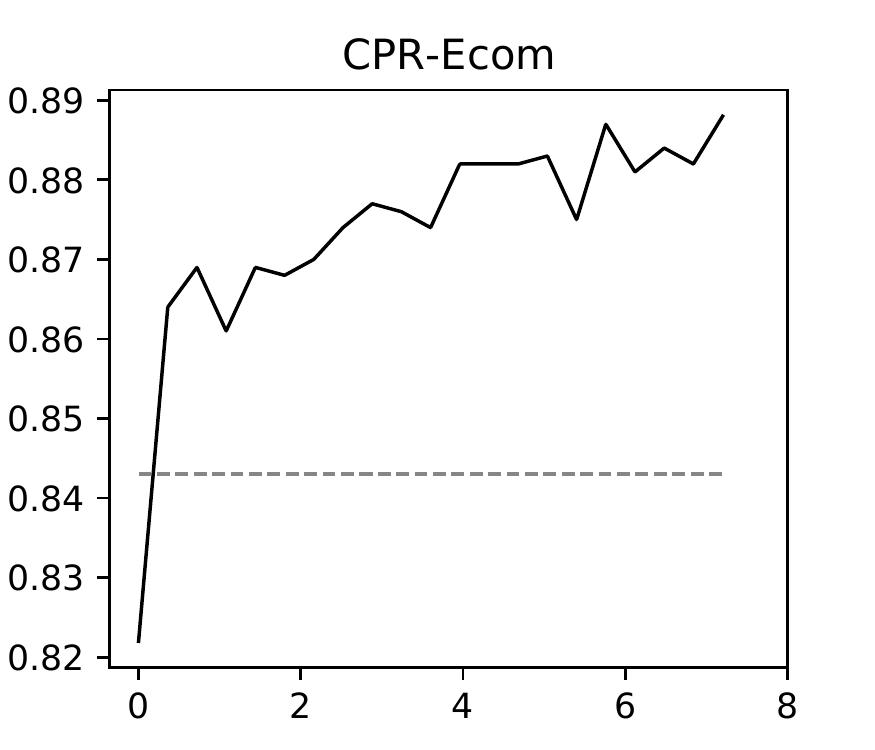}} 
    \subfloat{\includegraphics[width=0.33\linewidth]{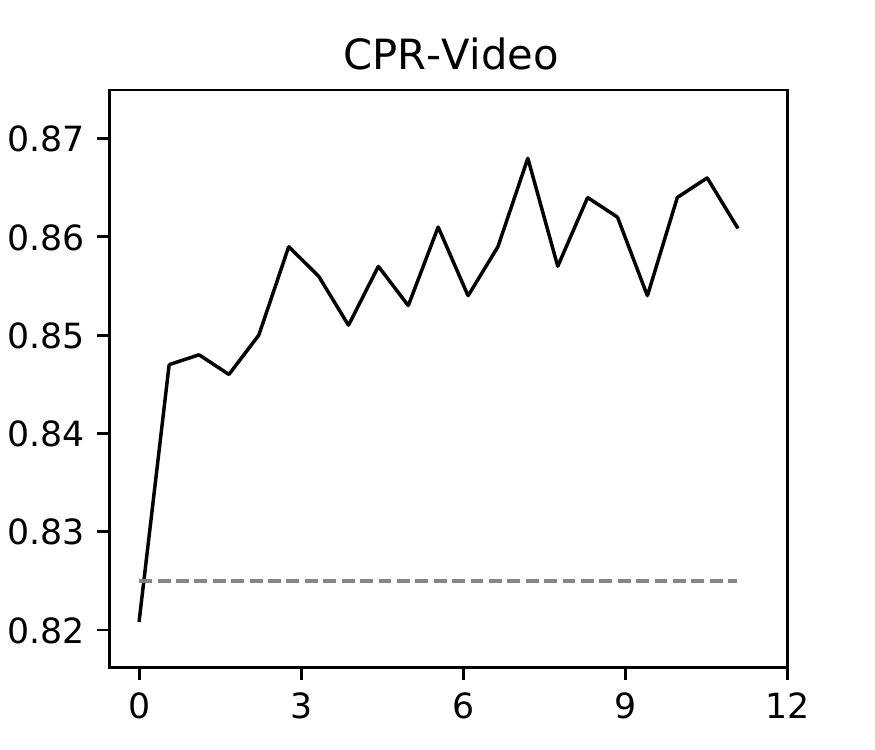}}      
    \subfloat{\includegraphics[width=0.33\linewidth]{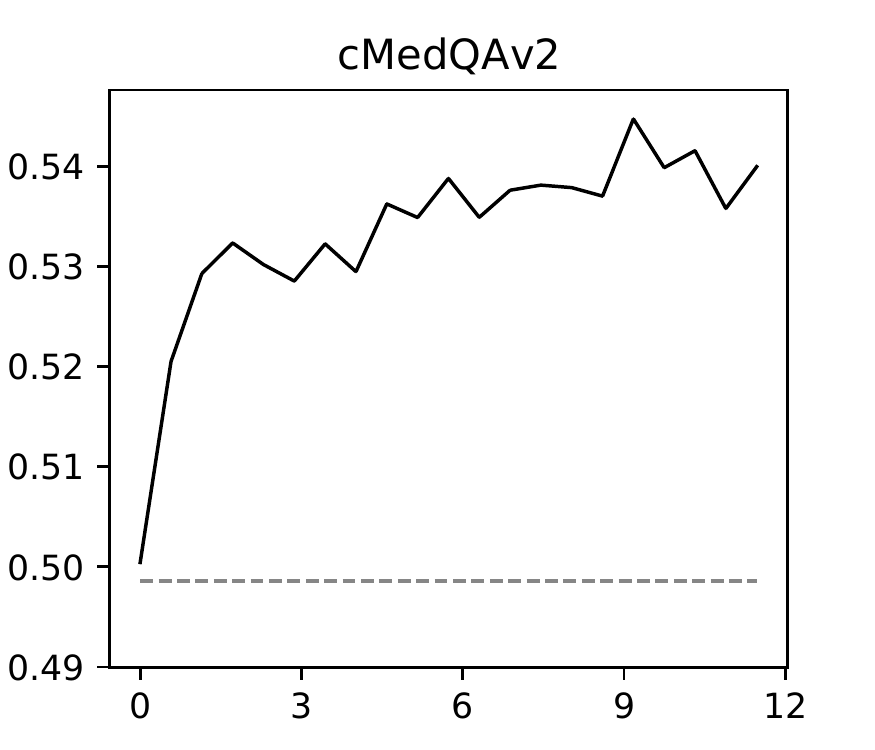}} 
    \caption{Impact of the target-domain training time~(in GPU hours) for DDR. We report R@1000 at different times when DAM is trained on the target-domain corpus. 
    The dashed line shows DR performance.
    \label{fig:perform_vs_steps}}
\end{figure}

The domain adaption capacity of DDR originates from the flexible DAM, which can be unsupervisedly updated based on the target-domain corpus. To investigate how it helps improve the out-of-domain performance of DDR, we illustrate the ranking performance (R@1000) at different times when we train DAM in the target domain in Figure~\ref{fig:perform_vs_steps}. DAM is trained for 50,000 steps on Lotte and for 20,000 steps on other datasets. The variance in training time is caused by different numbers of training steps and different document length distribution. The dashed line indicates the performance of DR. 
In the beginning, DR and DDR perform comparably because both models suffer from the domain shift. When we start to train DAM on the target-domain corpus, the performance of DDR quickly improves. It continuously improves after a long period of training especially on datasets with large corpora such as Lotte-Sci and Lotte-Tech. 
Therefore, the results highlight the significant and robust contribution of updating DAM on the target-domain corpus.

\subsubsection{Benefits of Disentangled Modeling} \mbox{}

\begin{table}
	\small
    \centering
    \caption{Benefits of disentangled modeling. We report R@1000 results. `en' and `zh' are abbreviations for English and Chinese, respectively. Results suggest DDR performs comparably to DDR$\backslash$D while being more training-efficient.}
    \label{tab:ablation_disentangled_modeling}
    \begin{threeparttable}
    \begin{tabular}{l|C{11mm}C{9mm}|C{11mm}C{9mm}}
    \toprule
        Training & \multicolumn{2}{c|}{Contrast} & \multicolumn{2}{c}{Distillation}  \\
        Models 				& DDR$\backslash$D &  DDR & DDR$\backslash$D &  DDR  \\
		\makecell[cl]{Finetuning Times} & $N$\tnote{$\dagger$} & $1$ & $N$\tnote{$\dagger$} & $1$ \\
         \midrule
        TREC-Covid~(en)		& 0.404  & 0.432  & 0.432  & \textbf{0.434}   \\ 
        Lotte-Write~(en) 	& 0.796  & \textbf{0.805}  & 0.784  & 0.791   \\ 
        Lotte-Rec~(en) 		& 0.859  & \textbf{0.866}  & 0.852  & 0.849   \\ 
        Lotte-Tech~(en) 	& \textbf{0.713}  & 0.711  & 0.695  & 0.698   \\ 
        Lotte-Life~(en) 	& \textbf{0.904}  & 0.903  & 0.897  & 0.899   \\ 
        Lotte-Sci~(en) 		& 0.513  & 0.520  & \textbf{0.534}  & 0.531   \\ 
        CPR-Ecom~(zh)		& 0.872  & \textbf{0.882}  & 0.864  & 0.861   \\ 
        CPR-Video~(zh)		& 0.853  & 0.859  & 0.879  & \textbf{0.887}   \\ 
        CPR-Medical~(zh) 	& \textbf{0.673}  & 0.668  & 0.636  & 0.643   \\ 
        cMedQAv2~(zh) 		& 0.530  & \textbf{0.532}  & 0.490  & 0.500   \\ 
    \bottomrule
    \end{tabular}
    \begin{tablenotes}
    	\footnotesize
    	\item [$\dagger$] N is the number of target domains. DDR$\backslash$D is finetuned per target domain, while DDR trains REM only once.
    \end{tablenotes}
    \end{threeparttable}
\end{table}

Disentangled modeling enables training one REM that can be shared across different target domains.
If we remove disentangled modeling, i.e., there is no REM or DAM but a Transformer model as a whole, we can still borrow the idea of DAM to unsupervisedly train Transformer in the target domain, but then we have to finetune the Transformer on supervised data to learn relevance estimation.
For example, we first unsupervisedly train a Transformer on the corpus of TREC-Covid with masked language modeling for domain adaption, then we finetune the model on MS MARCO to learn relevance estimation. Note that we have to finetune the model on MS MARCO because we assume no supervised data is available in the target domain.
Since in the first step we train different unsupervised Transformers for different domains, the second supervised learning process cannot be shared across domains and have to be executed for each target domain.
Therefore, although the model also acquires the domain modeling ability for the target domain like DDR, it is repeatedly finetuned on the same supervised data with different initializations. We call this method DDR$\backslash$D. 
In Table~\ref{tab:ablation_disentangled_modeling}, we show the recall performance of DDR and DDR$\backslash$D. We can see that DDR$\backslash$D performs similarly to DR.
Nevertheless, it induces substantial training costs due to repeated supervised learning when applied to each target domain, making it much less practical than DDR which only requires one-time supervised learning. 
Therefore, disentangled modeling considerably improves training efficiency while preserving ranking effectiveness.

\begin{table}
	\small
    \centering
    \caption{Ranking performance when the Relevance Estimation Module is implemented as LoRA, Parallel Adapter~(PA), and their combination. We report R@1000 results. `en' and `zh' are abbreviations for English and Chinese, respectively.}
    \label{tab:ablation_adapter}
    \begin{tabular}{l|C{9mm}C{9mm}|C{10mm}|C{16mm}}
    \toprule
    	Training & \multicolumn{4}{c}{Contrastive Learning} \\
      	Architecture & \multicolumn{2}{c|}{LoRA} & {PA} & LoRA \& PA \\ 
        BottleNeck size & 96 & 192 & 192 & 192 \& 192 \\ 
        \#Train Params 	& 3.20\% & 6.50\% & 3.20\% & 9.70\% \\ 
        \#Infer Params  & +0\% 	 & +0\%   & +3.20\% & +3.20\% \\ \midrule
        TREC-Covid~(en) & \textbf{0.432}  & 0.412  & 0.428  & 0.432  \\ 
        Lotte-Write~(en)& \textbf{0.809}  & 0.808  & 0.795  & 0.805  \\ 
        Lotte-Rec~(en)  & \textbf{0.873}  & 0.870  & 0.859  & 0.866  \\ 
        Lotte-Tech~(en) & \textbf{0.716}  & 0.715  & 0.694  & 0.711  \\ 
        Lotte-Life~(en) & \textbf{0.915}  & 0.913  & 0.898  & 0.903  \\ 
        Lotte-Sci~(en) 	& 0.519  & 0.512  & 0.500  & \textbf{0.520}  \\ 
        CPR-Ecom~(zh) 	& 0.866  & 0.868  & \textbf{0.882}  & \textbf{0.882}  \\ 
        CPR-Video~(zh) 	& 0.845  & 0.843  & 0.850  & \textbf{0.859}  \\ 
        CPR-Medical~(zh)& 0.661  & 0.659  & \textbf{0.668}  & \textbf{0.668}  \\ 
        cMedQAv2~(zh) 	& 0.533  & 0.532  & \textbf{0.537}  & 0.532  \\ 
    \bottomrule
    \end{tabular}
\end{table}

\subsubsection{Impact of REM Architecture} \mbox{}

After showing the benefits of framework designs, now we analyze the impact of our implementation.
This section presents how different implementations of REM influence effectiveness and efficiency.  
The default implementation of REM in this paper is the combination of two delta tuning modules, i.e., LoRA~\cite{hu2021lora} and PA~\cite{he2022towards}, but there could also be other implementations. In this section, we additionally use LoRA and PA separately as REM. The ranking performance~(R@1000) is shown in Table~\ref{tab:ablation_adapter}. To illustrate how REM affects training and inference efficiency, Table~\ref{tab:ablation_adapter} also presents the percentage of trainable parameters during supervised learning and the percentage of extra parameters during inference. Delta tuning methods utilize frozen backbone and only update a small number of their parameters. Therefore, they can improve training efficiency. During inference, LoRA can merge itself into the backbone and causes zero overhead, whereas PA requires extra computation but the overhead is still marginal thanks to the small number of parameters.  
As for ranking effectiveness, different architectures indeed result in slightly different results. LoRA is more effective on English datasets, 
while PA is more effective on Chinese datasets. Our default LoRA \& PA implementation seems like a trade-off between the two methods. The intriguing phenomenon seems to suggest different languages may have different patterns and thus require different model structures. It also implies that the model performance can be further improved with a stronger model architecture. We leave these to future studies.

\subsubsection{Effectiveness of Optimization Designs} \mbox{}
\label{sec:ablation_implementation_df_si}

\begin{table}
	\small
    \centering
    \caption{Impact of disentangled finetuning~(DF) and sequential initialization~(SI). We report R@1000 results. `en' and `zh' are abbreviations for English and Chinese, respectively.}
    \label{tab:ablation_training_df_si}
    \begin{threeparttable}
    \begin{tabular}{l|C{12mm}C{12mm}C{11mm}}
    \toprule
        Training  & \multicolumn{3}{c}{Contrastive Learning} \\ 
        Models & DDR$\backslash$DF~\tnote{a} & DDR$\backslash$SI~\tnote{b} &  DDR  \\ \midrule
        TREC-Covid~(en) & 0.417  & 0.362  & \textbf{0.432}   \\ 
        Lotte-Write~(en)& 0.792  & 0.788  & \textbf{0.805}     \\ 
        Lotte-Rec~(en) 	& 0.818  & 0.840  & \textbf{0.866}     \\ 
        Lotte-Tech~(en) & 0.665  & 0.690  & \textbf{0.711}     \\ 
        Lotte-Life~(en) & 0.869  & 0.874  & \textbf{0.903}     \\ 
        Lotte-Sci~(en) 	& 0.463  & 0.480  & \textbf{0.520}     \\ 
        CPR-Ecom~(zh) 	& 0.871  & 0.880  & \textbf{0.882}     \\ 
        CPR-Video~(zh) 	& 0.846  & 0.845  & \textbf{0.859}     \\ 
        CPR-Medical~(zh)& 0.626  & 0.663  & \textbf{0.668}     \\ 
        cMedQAv2~(zh)	& 0.515  & 0.530  & \textbf{0.532}    \\ 
    \bottomrule
    \end{tabular}
    \begin{tablenotes}
    	\footnotesize
    	\item [a] DF is short for disentangled finetuning. 
    	\item [b] SI is short for sequential initialization. 
    \end{tablenotes}
    \end{threeparttable}
\end{table}


Besides model architectures, the other important part of our implementation is optimization designs tailored for DDR. DDR employs two particular training designs, i.e., disentangled finetuning~(DF) and sequential initialization~(SI). The former aims to train a domain-invariant REM, and the latter aims to improve the compatibility between DAM and REM. Now we investigate their efficacy. We provide two variants, DDR$\backslash$DF and DDR$\backslash$SI, that remove the two designs, respectively. Concretely, DDR$\backslash$DF does not adapt the source DAM in the training corpus and replaces the source DAM with the BERT checkpoint. DDR$\backslash$SI initializes the target DAM with the BERT checkpoint instead of the source DAM. We report R@1000 performance in Table~\ref{tab:ablation_training_df_si}. According to the ablation results, removing either of them results in performance loss. Removing disentangled finetuning makes REM incapable of generalizing to new domains because it fits the domain features of the training data. Removing sequential initialization causes incompatibility between REM and DAM. 
Therefore, our proposed techniques effectively address the optimization challenges brought by disentangled modeling.

\section{Conclusion and Future Work}

In this paper, we propose a novel DR framework named Disentangled Dense Retrieval~(DDR), which supports effective and flexible domain adaption. DDR consists of a relevance estimation module~(REM) and a domain adaption module~(DAM). REM learns domain-invariant matching patterns with supervised data only once and DAM learns domain-specific features unsupervisedly to mitigate the domain shift. 
Three designs are employed to effectively implement DDR. 
Firstly, we propose an architecture that effectively disentangles the two modules and induces marginal overhead. 
Secondly, we train REM with a disentangled finetuning technique to help it generalize to various domains.
Thirdly, we specially initialize DAM to improve its compatibility with REM.
Comprehensive experiments show prominent advantages of DDR compared to DR and traditional retrieval methods.

We believe disentangled modeling is a promising area for neural ranking. It is proposed and implemented to improve the adaption capacity of DR in this paper.  
It can be extended to other ranking infrastructures beyond DR, such as neural sparse retrieval~\cite{lin2021few, mallia2021learning, formal2021splade}, late-interaction models~\cite{Khattab2020ColBERTEA, macavaney2020efficient}, and interaction-based models~\cite{nogueira2019passage}, and neural document expansion models~\cite{nogueira2019doc2query, nogueira2019document}. 
Its effectiveness can also be further improved with a better implementation design, e.g., a stronger REM architecture or a better training objective for DAM. 
Our work sheds light on this promising direction, and we leave these explorations to future work.


\bibliographystyle{ACM-Reference-Format}
\balance
\bibliography{references}

%
%
%

\end{document}